\documentclass{article}

\usepackage{arxiv}

\usepackage[utf8]{inputenc} 
\usepackage[T1]{fontenc}    
\usepackage{hyperref}       
\usepackage{url}            
\usepackage{booktabs}       
\usepackage{amsfonts}       
\usepackage{nicefrac}       
\usepackage{microtype}      
\usepackage{graphicx}

\hypersetup{
    colorlinks=true,   
    linkcolor=blue,   
    citecolor=blue,   
    urlcolor=blue      
}

\title{Learning-to-Rank with BERT in TF-Ranking}

\author{
  Shuguang Han, Xuanhui Wang, Michael Bendersky and Marc Najork\\[6pt]
  TF-Ranking Team, Google Research, Mountain View, CA \\[3pt]
  \texttt{\{hanshuguang,xuanhui,bemike,najork\}@google.com}
}

\begin{document}
\maketitle

\begin{abstract}
This paper describes a machine learning algorithm for document (re)ranking, in which queries and documents are firstly encoded using BERT~\cite{devlin2018bert}, and on top of that a learning-to-rank (LTR) model constructed with TF-Ranking (TFR)~\cite{pasumarthi2019tf} is applied to further optimize the ranking performance. This approach is proved to be effective in a public MS MARCO benchmark~\cite{bajaj2016ms}. Our first two submissions achieve the best performance for the \emph{passage re-ranking} task~\cite{han2020ensemble}, and the second best performance for the \emph{passage full-ranking} task as of April 10, 2020~\cite{han2020fullrankingensemble}. To leverage the lately development of pre-trained language models, we recently integrate RoBERTa~\cite{liu2019roberta} and ELECTRA~\cite{clark2020electra}. Our latest submissions improve our previously state-of-the-art re-ranking performance by 4.3\%~\cite{han2020reranking-multibert}, and achieve the third best performance for the full-ranking task~\cite{han2020fullranking-multibert} as of June 8, 2020. Both of them demonstrate the effectiveness of combining ranking losses with BERT representations for document ranking.
\end{abstract}

\section{Introduction}
Recently, neural network models built on top of pretrained language models such as BERT~\cite{devlin2018bert} have achieved state-of-the-art performance on various machine learning tasks including question answering~\cite{lan2019albert}, key-phrase extraction \cite{xiong2019open}, as well as document and passage ranking~\cite{nogueira2019multistage, nogueira2019passage}. In this paper, we are focusing on passage ranking, and particularly the MS MARCO passage full ranking and re-ranking tasks~\cite{bajaj2016ms}.

A common way to incorporate BERT for ranking tasks is to construct a finetuning classification model with the goal of determining whether or not a document is relevant to a query~\cite{nogueira2019passage}. The resulting predictions are then used for ranking documents. We argue that such an approach is less suited for a ranking task, compared to a pairwise or listwise learning-to-rank (LTR) algorithm, which learns to distinguish relevance for document pairs or to optimize the document list as a whole, respectively~\cite{liu2009learning}.

To this end, we propose \textbf{TFR-BERT}, a generic document ranking framework that builds a LTR model through finetuning BERT representations of query-document pairs within TF-Ranking\footnote{TF-Ranking official page: \url{https://github.com/tensorflow/ranking}}. We apply this approach on the MS MACRO benchmark, and our submissions achieve the best leaderboard performance for the passage re-ranking task~\cite{han2020reranking-multibert}, and the third best performance for the passage full ranking task~\cite{han2020fullranking-multibert}. This demonstrates the effectiveness of combining ranking losses with BERT representations for passage ranking.

\section{TFR-BERT}
\begin{figure}[!tbp]
\centerline{\includegraphics[scale=0.40]{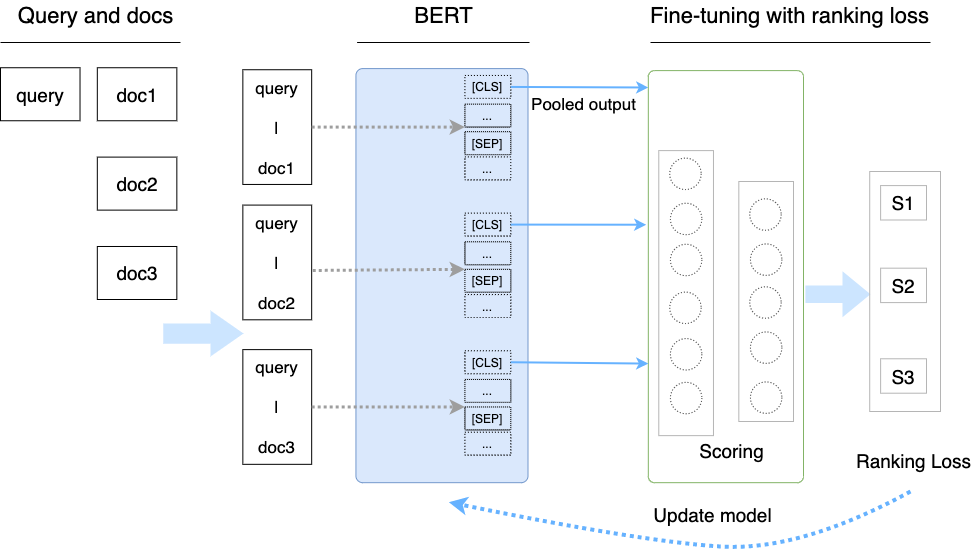}}
\caption{An illustration of the TFR-BERT framework, in which a Learning-to-Rank model is constructed on top of the BERT representations of query-document pairs.}
\label{fig:bert-ranking}
\end{figure}

Our TFR-BERT model can be illustrated by Figure \ref{fig:bert-ranking}. Documents (passages) for a given query will be firstly flattened to query-document (query-passage) pairs, and then passed through BERT layers\footnote{We use BERT checkpoints downloaded from the official BERT page: \url{https://github.com/google-research/bert}.}. Specifically, query and each document (passage) are treated as two sentences, and are further concatenated to the following format:
\vspace{0.5em}

\centerline{\emph{[CLS] query text [SEP] passage text [SEP]}}

Here, [CLS] indicates the start of a sequence and [SEP] denotes a separator between the first and second sentences. We also truncate the passage text if the whole sequence exceeds the maximum length of 512 tokens.

After that, the pooled BERT outputs (i.e., the hidden units of the [CLS] token) are fed into a ranking model built from TF-Ranking~\cite{pasumarthi2019tf}. TF-Ranking provides a variety of pointwise, pairwise and listwise losses, which enable us to compare different  LTR approaches in our TFR-BERT model. 

\section{MS MARCO Experiment}
To understand the performance of TFR-BERT, we conduct a set of experiments using the publicly available MS MARCO dataset. The dataset contains 1 million real Bing queries (each query is a question), and 8.8 million candidate documents (each document is a passage). For each query, it also provides zero or more respective relevant passages marked by human annotators. In this work, we study both the passage \emph{full ranking} and \emph{re-ranking} tasks.

\textbf{Passage Re-ranking Task}. For each query, we are given the top 1000 candidate passages retrieved by BM25. The goal is to re-rank passages by their relevance to the query, i.e. the likelihood to be an answering passage for the question.

\textbf{Passage Full Ranking Task}. While the re-ranking performance is bounded by the recall of top 1000 passages from BM25, in this full ranking task, we are asked to rank relevant documents for each query from the whole collection of 8.8 million passages.

\textbf{Ranking Dataset}. To create the training set, we employ the data from \textit{triples.train.full.tsv}. In this file, each data record is a triple containing the content of a query, a relevant passage and an irrelevant passage (query and the relevant passage can repeat multiple times in the dataset). For each query, there are roughly 1000 passages, and (in most cases) only one of them is relevant\footnote{More details about this dataset can be found in https://github.com/nyu-dl/dl4marco-bert.}.

To better support the pairwise and listwise ranking models, we further group triples by query. Therefore, we obtain a list of up to 1000 passages for each query. With regards to the  computer memory limit, we further break this passage list into roughly 90 lists, each taking one relevant passage and 11 irrelevant passages; thereby, creating a set of passage lists with size up to 12. Note that the above process is only used when building the training data. We leave the \emph{dev} and \emph{eval} datasets intact -- 1000 passages per each query are present for these datasets.

\textbf{Training}. Our models are trained on TPU V3. We set the list size to be 12, as described above. The batch size is set to 32. As a result, a number of 32 * 12 = 384 query-document pairs are used in each training step. We checkpoint each model at the 50K steps. Our ensemble approach, which will be introduced in Section \ref{subsec::reranking}, aggregates over multiple models, each following the above training process. 

\section{Experimental Results}
In this section, we report the results obtained by TFR-BERT, in which we take into account all of the pointwise, pairwise and listwise ranking approaches. The ranking models are constructed using the open-source TF-Ranking code. For more details about their implementation, the readers may refer to Pasumarthi et al. \cite{pasumarthi2019tf} and Bruch et al.~\cite{Bruch+al:2019}.

\subsection{Our Submissions}
We made five submissions to the MS MARCO leaderboard (\url{https://microsoft.github.io/msmarco/}), as listed below. Submission \#1, \#2 and \#4 focused on the passage re-ranking task (Section \ref{subsec::reranking}), whereas the other two submissions addressed the passage full ranking task (Section \ref{subsec::fullranking}).

For pre-trained language models, we used the \textit{BERT-Large, Uncased} checkpoint \cite{devlin2018bert} for submissions \#1 to \#3. Later on, we switched to the \textit{BERT-Large, Uncased (Whole Word Masking)} checkpoint for submissions \#4 and \#5 because of its better performance. For RoBERTa, we adopted the \textit{roberta.large} checkpoint. And for ELECTRA, we utilized the \textit{ELECTRA-Large} checkpoint.

More specifically, Submission \#1 was a single run of TFR-BERT with softmax loss; Submission \#2 was an ensemble of pointwise, pairwise and listwise TFR-BERT models; Submission \#3 adopted the same ensemble technique as Submission \#2, but re-ranked top 1000 passages from both BM25 and DeepCT~\cite{dai2019context}, and further combined the two ranking lists; Submission \#4 only adopted the listwise loss in TF-Ranking but used ensemble over BERT, RoBERTa and ELECTRA; Submission \#5 applied the same ensemble technique as Submission \#4, but combined both DeepCT~\cite{dai2019context} and BM25 results for re-ranking.

\begin{itemize} 
\item \textbf{Submission \#1} (re-ranking): TF-Ranking + BERT (Softmax Loss, List size 6, 200k steps)~\cite{han2020softmax}.

\item \textbf{Submission \#2} (re-ranking): TF-Ranking + BERT (Ensemble of pointwise, pairwise and listwise losses)~\cite{han2020ensemble}.

\item \textbf{Submission \#3} (full ranking): DeepCT Retrieval + TF-Ranking BERT Ensemble~\cite{han2020fullrankingensemble}.

\item \textbf{Submission \#4} (re-ranking): TF-Ranking Ensemble of BERT, RoBERTa and ELECTRA~\cite{han2020reranking-multibert}.

\item \textbf{Submission \#5} (full ranking): DeepCT + TF-Ranking Ensemble of BERT, RoBERTa and ELECTRA~\cite{han2020fullranking-multibert}.
\end{itemize}

\subsection{Re-ranking Experiments} \label{subsec::reranking}
Experimental results for re-ranking tasks are provided in Table \ref{tbl::reranking-performance}. In addition to the official BM25 and Duet V2 baselines, we also include a baseline from Nogueira and Cho \cite{nogueira2019passage}. 

\begin{table}[!htbp]
\fontsize{10}{10}\selectfont
\caption{MRR@10 performance for passage \textbf{re-ranking}. Note that 1) only the models submitted to the leaderboard have MRR@10 for the Eval dataset, 2) for multiple BERT ensemble, we switched the checkpoint from \textit{BERT-Large, Uncased} to \textit{BERT-Large, Uncased (Whole Word Masking)}, which slightly improved MRR@10 from 0.3856 to 0.3898.}
\begin{center}
\renewcommand{\arraystretch}{1.7}
\begin{tabular}{ll|c|c}
\hline
           & Model                                  & Dev (MRR@10)  & Eval (MRR@10) \\ \hline
Baselines  & BM25                                   & 0.1670          & 0.1649        \\ 
           & Duet V2 (\cite{mitra2019updated})      & 0.2517          & 0.2527        \\
           & BERT + Small training (\cite{nogueira2019passage}) & 0.3653  & 0.3587    \\
           & Previous Leaderboard Best~\cite{mingyan2019}                   & 0.3730      & 0.3676  \\ \hline
TFR-BERT Single Run   & Sigmoid cross entropy loss (pointwise)     & 0.3716      & -     \\
           & Pairwise logistic loss (pairwise)          & 0.3718      & -     \\
           & Softmax loss (listwise)                    & 0.3725      & - \\
           & \textbf{Submission \#1}~\cite{han2020softmax}        & \textit{0.3782}      & \textit{0.3660} \\ \hline
Multiple Losses (Ensemble) & Sigmoid cross entropy loss (5 runs) & 0.3839      & -   \\
                  & Pairwise logistic loss (5 runs)     & 0.3849      & -   \\
                  & Softmax loss (5 runs)               & 0.3856      & -   \\
                  & \textbf{Submission \#2}~\cite{han2020ensemble} & \textit{0.3877}    & \textit{0.3747}   \\\hline
Multiple BERTs (Ensemble) & BERT (5 runs, listwise loss*)  & 0.3898      & -   \\
                  & RoBERTa (5 runs, listwise loss)      & 0.3958      & -   \\
                  & ELECTRA (5 runs, listwise loss)               & 0.3976      & -   \\
                  & \textbf{Submission \#4}~\cite{han2020reranking-multibert} & \textbf{0.4046}    & \textbf{0.3905}   \\\hline
\end{tabular}
\end{center}
\label{tbl::reranking-performance}
\end{table}

\textbf{TFR-BERT Single Run.} We experimented with three types of TFR-BERT models -- pointwise model with sigmoid cross-entropy loss, pairwise model with pairwise logistic loss and listwise with softmax loss. We run each model 5 times, and the reported numbers are the average of 5 runs. For \textbf{Submission \#1}~\cite{han2020softmax}, we choose the softmax loss run with the best MRR@10 performance on the Dev data set over the 5 runs. 

According to Table \ref{tbl::reranking-performance}, TFR-BERT models outperform the official baselines by a large margin. More importantly, they further improve upon the existing state-of-the-art approach \cite{nogueira2019passage} that uses the same training data and BERT checkpoint. This demonstrates the effectiveness of combining ranking losses with BERT representations for passage ranking. 

The \textbf{Submission \#1} achieved the second best performance for the passage re-ranking task at the time of its submission on March 19, 2020. Compared with the best method at that time~\cite{mingyan2019}, which used auxiliary information to enrich BERT, and introduced additional index information for ranking\footnote{However, the author did not disclose further details about his approach.}, TFR-BERT only adopted the original BERT checkpoint, and can be reproduced easily in TF-Ranking.

\textbf{Ensemble of Multiple Losses.} After a manual examination of model predictions, we discovered that, despite similar MRR performance, different TFR-BERT runs (even with the same type of loss) show non-trivial difference in predictions. Therefore, we further include an approach to \textbf{ensemble} models trained from different runs. It worked as follows:

\begin{itemize} 
\item[1:] Supposes we have $n$ runs (models) to ensemble: $R_1$, $R_2$, ..., $R_n$.

\item[2:] For each run $R_k$ and a query $q_i$, we rank the corresponding documents based on prediction scores, and then obtain the rank position $P_{k, i,j}$ for each document $d_j$.

\item[3:] For each query $q_i$, we re-compute a new score $s_{i,j}$ for document $d_j$ based on the average reciprocal rank ($\frac{1}{n} \sum_k{\frac{1}{P_{k,i,j}}}$) of $n$ runs. 

\item[4:] Finally, we rank documents based on the new score $s_{i,j}$.
\end{itemize}

We firstly experimented with the ensemble of 5 different runs using the same loss function.  According to Table \ref{tbl::reranking-performance}, the ensemble approach improves the performance of a single run by 3.5\% for all three loss functions. Through a further ensemble over all the three loss functions (total of 15 runs), we achieve the best overall MRR on the Dev data set. The 15-run ensemble is chosen as the \textbf{Submission \#2} \cite{han2020ensemble}, which outperforms the previously best submission~\cite{mingyan2019} by 4.0\% on the development dataset, and 1.9\% on the evaluation dataset.

\textbf{Ensemble of Multiple BERTs.} To incorporate the recent advancement of pre-trained BERT models, we further integrated RoBERTa~\cite{liu2019roberta} and ELECTRA~\cite{clark2020electra} into TF-Ranking. The ensemble process for each BERT model worked the same as the above. From Table \ref{tbl::reranking-performance}, we observed that ensemble with RoBERTa slightly outperformed BERT, and ensemble with ELECTRA slightly outperformed RoBERTa. Through a further ensemble over all of the three models (total of 15 runs, \textbf{Submission \#4}), we achieve the best MRR@10 for the re-ranking task \cite{han2020reranking-multibert}, outperforming the previously best performance (also from us~\cite{han2020ensemble}) by 4.4\% on the dev dataset and by 4.3\% on the evaluation dataset.

\subsection{Full Ranking Experiments} \label{subsec::fullranking}
In addition to the re-ranking task, we made another submission to the full ranking task, in which we re-ranked the top 1000 passages from both BM25 and DeepCT~\cite{dai2019context} using the TFR-BERT ensemble model, and further combined the two resulting ranking lists. It worked as follows.

\begin{itemize} 
\item[1:] Re-rank the top 1000 passages retrieved by BM25 using the TFR-BERT ensemble model.

\item[2:] Re-rank the top 1000 passages retrieved by DeepCT~\cite{dai2019context} using the TFR-BERT ensemble model.

\item[3:] Combine the re-ranking scores (we use reciprocal rank to be consistent with the ensemble model) from the above two lists. For passages occurring in both lists, we take the average; otherwise, we keep its original score.  

\item[4:] Finally, we re-rank passages based on the new score.
\end{itemize}

The full-ranking results are reported in Table \ref{tbl::full-ranking-performance}. Same as the re-ranking results, we include the official BM25 and Duet V2 baselines for reference. In addition, we introduce a baseline (W-index + BERT F-rerank) from Dai et al. \cite{dai2019context}, as it is the original entry that proposed the DeepCT retrieval approach. 

According to Table \ref{tbl::full-ranking-performance}, we discovered that DeepCT helps boost the re-ranking of BM25 results by a large margin, and a further combination of both BM25 and DeepCT re-ranked lists brings additional gains. With Submission \#3, we achieved the second best overall performance on the leaderboard as of April 10, 2020. With the recent Submission \#5, we further improved our previous performance, and obtained the third best performance on the leaderboard as of June 8, 2020 (with tens of new leaderboard submissions in between).

The above results, again, demonstrate the effectiveness and robustness of the TFR-BERT ensemble model -- it works well on both re-ranking and full ranking tasks, and more importantly, it does not require auxiliary information other than the original BERT checkpoints, and can be easily reproduced with TF-Ranking.

\begin{table}[!htbp]
\fontsize{10}{10}\selectfont
\caption{MRR@10 performance for the passage \textbf{full ranking} task. Note that only the models submitted to the leaderboard have MRR@10 for the Eval dataset.}
\begin{center}
\renewcommand{\arraystretch}{1.6}
\begin{tabular}{ll|c|c}
\hline
           & Model                                  & Dev (MRR@10)  & Eval (MRR@10) \\ \hline
Baselines  & BM25                                   & 0.1670          & 0.1649        \\ 
           & Duet V2 (\cite{mitra2019updated})      & 0.2517          & 0.2527        \\
           & W-index + BERT-F rerank (\cite{dai2019context}) & 0.3935  & 0.3877    \\
           & Leaderboard Best~\cite{sun2020} (as of April 10, 2020)            & 0.4012      & 0.3998  \\
           & Leaderboard Best~\cite{sun2020-drbert} (as of June 8, 2020)            & 0.4200      & \textbf{0.4190}  \\ \hline
Multiple Losses (Ensemble) & Re-ranking over BM25   & 0.3877      &  0.3747  \\
                  & Re-ranking over DeepCT & 0.4012      & -   \\
                  & \textbf{Submission \#3}: combining the above~\cite{han2020fullrankingensemble} & 0.4049 & 0.3946   \\\hline
Multiple BERTs (Ensemble) & Re-ranking over BM25   & 0.4046      &  0.3905  \\
                  & Re-ranking over DeepCT & 0.4175      & -   \\
                  & \textbf{Submission \#5}: combining the above~\cite{han2020fullranking-multibert} & \textbf{0.4213} & 0.4073   \\\hline
\end{tabular}
\end{center}
\label{tbl::full-ranking-performance}
\end{table}

\section{Conclusion}
In this paper, we propose the TFR-BERT framework for document and passage ranking. It combines state-of-the-art developments from both pretrained language models, such as BERT, and learning-to-rank approaches. Our experiments on the MS MARCO passage ranking task demonstrate its effectiveness.

\section{Acknowledgement}
We would like to thank Zhuyun Dai from Carnegie Mellon University for kindly sharing her DeepCT retrieval results. We would also like to thank Sebastian N. Bruch from Google Research for creating the MS MARCO datasets for our experiments. This work would not be possible without the support provided by the TF-Ranking team.

\bibliographystyle{unsrt}  
\bibliography{references} 

\end{document}